\documentclass{aa_gaia}  
\usepackage{hyperref}

\usepackage{xcolor}
\definecolor{dark-red}{rgb}{0.9,0.0,0.0}
\definecolor{dark-blue}{rgb}{0.15,0.15,0.9}
\definecolor{dark-green}{rgb}{0.15,0.8,0.15}
\definecolor{medium-blue}{rgb}{0,0,0.9}
\hypersetup{
    colorlinks, linkcolor=red,
    citecolor={dark-blue} , urlcolor={medium-blue}
}
\usepackage{changepage} 

\usepackage{graphicx}
\usepackage{txfonts}
\usepackage{soul}
\usepackage{relsize}
\begin{document}

   \title{TESS exoplanet candidates validated with HARPS archival data}

   \subtitle{A massive Neptune around GJ\,143 and two Neptunes around HD\,23472\thanks{Based on observations collected at 
the European Organization for Astronomical Research in the Southern Hemisphere under ESO 
programmes 072.C-0488, 183.C-0972, 085.C-0019, 087.C-0831, 089.C-0732, 090.C-0421} \thanks{Tables with the reprocessed HARPS time series
are only available in electronic form at the CDS via anonymous ftp to cdsarc.u-strasbg.fr (130.79.128.5)
or via http://cdsarc.u-strasbg.fr/viz-bin/qcat?J/A+A/622/L7} }

    \author{Trifon\,Trifonov\inst{1} \and Jan\,Rybizki\inst{1} \and Martin\,K\"urster\inst{1}
}
   \institute{Max-Planck-Institut f\"ur Astronomie,
              K\"onigstuhl 17, D-69117 Heidelberg, Germany\\
              \email{trifonov@mpia.de} 
              }

   \date{Received 10 December 2018 / Accepted 14 January 2019}
 

  \abstract
   {}   
   {
 We aim at the discovery of new planetary systems by exploiting the transit light-curve results from observations made in TESS orbital observatory Sectors 1 and 2 
  and validating them with precise Doppler measurements obtained from archival HARPS data.
   }
   {
    Taking advantage of the reported TESS transit events around GJ\,143 (TOI 186) and HD\,23472 (TOI 174), we 
    modeled their HARPS precise Doppler measurements and derived orbital parameters for these two systems.
   }
   {
     For the GJ\,143 system, TESS has reported only a single transit, and thus its period is unconstrained from photometry.
     Our radial velocity analysis of GJ\,143 reveals the full Keplerian solution of the system, which is consistent with 
      an eccentric planet with a mass almost twice that of Neptune and a period of $P_{\rm b}$ = $35.59_{-0.01}^{+0.01}$ days.
      Our estimates of the GJ\,143 b planet are fully consistent with the transit timing from TESS.
      We confirm the two-planet system around HD\,23472, which according to our analysis is composed of two Neptune-mass planets in a possible 5:3 mean motion resonance. 
   }
   {}


   \authorrunning{Trifonov et al.}
   \titlerunning{HARPS archival data confirmation of TESS targets}

   \maketitle


\section{Introduction}
 
The Transiting Exoplanet Survey Satellite \citep[TESS;][]{Ricker2015} has begun its planet hunt. 
As of December 2018, the official TESS data release  constitutes 
a total of 54 days of TESS observations from Sectors 1 and 2
taken between 22 July and 21 September.
Since then, a number of TESS planet candidates have been confirmed 
through ground-based Doppler spectroscopy \citep{Huang2018, Gandolfi2018, Wang2018, Jones2018},
and given this high rate of planet detections, a plethora of new planet discoveries during the 2\,yr-mission can easily be predicted.
In this paper we report the Doppler validation of two additional TESS systems
using archival HARPS spectra. 
We find an eccentric planet with almost two Neptune masses around the K dwarf GJ\,143 (TOI 186).
Its period of $P_{\rm b}$ = $35.59_{-0.01}^{+0.01}$\,d matches 
the single transit-timing event from TESS.
We provide evidence of the existence of a two-planet system around HD\,23472 (TOI 174) 
that is likely composed of two Neptune-mass planets in a period ratio of 5:3.

This paper is organized as follows: 
in Section \ref{Sec2} we present estimates of the stellar parameters of GJ\,143 and HD\,23472,
we present the available HARPS data and our approach to modeling the radial velocity (RV) data.
We present our results for these two exoplanet systems in Section \ref{Sec3}, and
in Section \ref{Sec4} we provide a brief summary and conclusions from this study.

\begin{table}[htp]

\caption{Stellar parameters with 1$\sigma$ uncertainties for the planet hosts.}
\label{table:phys_param}    


\centering          
\begin{tabular}{ l l l }     
\hline\hline  \noalign{\vskip 0.9mm}        
  Parameters\hspace{25 mm}   & GJ\,143   &  HD\,23472\\  
\hline    \noalign{\vskip 0.5mm}          


   Mass    [$M_{\odot}$]                    & 0.76$_{-0.02}^{+0.03}$    & 0.75$_{-0.03}^{+0.04}$ \\ \noalign{\vskip 0.5mm} 
  Radius    [$R_{\odot}$]                  & 0.73$\pm0.01$    & 0.73$\pm0.01$  \\ \noalign{\vskip 0.5mm} 
  Luminosity    [$L{_\odot}$]              & 0.197$\pm0.003$     &  0.231$\pm0.005$ \\ \noalign{\vskip 0.5mm}

\hline \noalign{\vskip 0.5mm}   

\end{tabular}

\end{table}

\section{GJ\,143 and HD\,23472 planet host candidates}
\label{Sec2}
We derived stellar parameters using the \texttt{isochrones} package \citep{Morton2015} 
together with the MIST stellar evolutionary tracks \citep{Dotter2016}, assuming zero 
extinction. Input data were the BVJK bands taken from Simbad together with the 
Gaia DR2 parallaxes (which we corrected for the zero-point offset). 
The inferred parameters and their precision for the two K-dwarfs GJ\,143
($\approx$ 16\,pc) and HD\,23472 ($\approx$ 39\,pc) are listed in Table\,\ref{table:phys_param}. 
We neglected systematic uncertainties, but a test using a different isochrone set yielded very similar results.

\subsection{HARPS data}
\label{Sec2a}
  
The HARPS spectra of GJ\,143 and HD\,23472 were reanalyzed
by our team prior to the TESS observations with the ultimate goal
to reanalyze all publicly available ESO HARPS spectra in order to
derive uniformly processed HARPS-Doppler measurements
and study possible small but significant instrument-related RV systematics \citep[i.e.,][]{TalOr2018}.
We plan to make these data publicly available as a service to the exoplanet community (Trifonov et al. in prep.).
All spectra in our MPIA HARPS-archive are reprocessed with the
 SpEctrum Radial Velocity AnaLyser 
\citep[SERVAL,][]{Zechmeister2017}
pipeline, which has been demonstrated to produce more precise RV measurements from HARPS spectra
than the standard ESO HARPS-DRS pipeline \citep[e.g., see][]{Trifonov2018a,Kaminski2018}.
In addition to the precise Doppler measurements, we inspected the 
activity index measurements of the H$_\alpha$ line, the differential line width of the spectral lines,
and the RV chromaticity (wavelength dependence), for more details see \citet[][]{Zechmeister2017}.

GJ\,143 has 54 HARPS measurements with a mean RV precision of 0.55 m\,s$^{-1}$ taken
 between November 2003 and December 2009, plus   
four measurements with a mean RV precision of 1.26 m\,s$^{-1}$ taken after the HARPS fiber 
upgrade \citep[since May 2015, see][]{LoCurto2015} in the nights from 18 to 26 December 2016. 
It is well known that after this intervention, 
HARPS is effectively a new instrument with a notable RV offset 
between the data taken before and after the fiber exchange.
\citet[][]{LoCurto2015} have estimated that for a star of spectral type K such as GJ\,143, 
the mutual RV offset is likely about 10$-$12 m\,s$^{-1}$,
which makes the detection of low-amplitude RV signals challenging.
We therefore decided to use the full HARPS data set of GJ\,143, but 
as a standard practice in this case, we modeled a mutual RV offset between the pre- and post-upgrade RVs.

HD\,23472 has only 14 precise HARPS RV measurements, which were taken between February 2004 and February 2013 (before the fiber upgrade).
The mean precision of the RV data is 1.37 m\,s$^{-1}$, but the RV scatter is $\approx 13.5$ m\,s$^{-1}$,
which can be taken as an indication that planetary companions are present. 
Of course, scatter like this is common in stars with no obvious planet detections, 
and the 14 RVs are not sufficient for a period search. 
Only after taking into account the {\em \textup{a priori}} knowledge from the TESS transit periods and 
orbital phases is it possible to model these data to search for the presence of planetary companions.

\begin{figure}[tp]
\begin{center}$
\begin{array}{cc} 

\includegraphics[width=9cm]{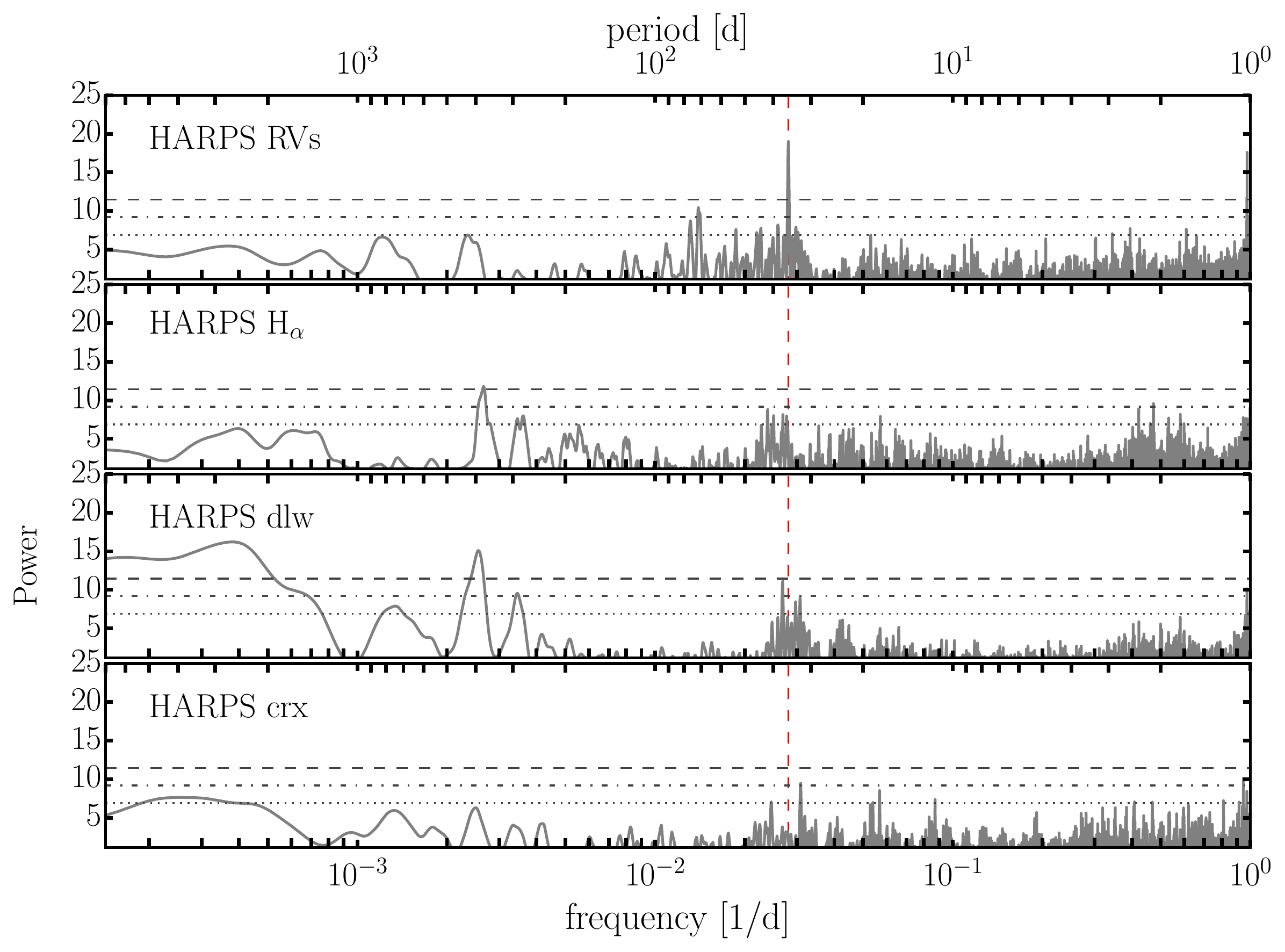} \\

\end{array} $
\end{center}

\caption{Top: GLS periodogram of the SERVAL time series of GJ\,143 with levels of the false alarm probability (FAP) 
of 10\% (dotted line), 1\% (dot-dashed line) and 0.1\% (dashed line).  Second to fourth panel:  Activity
indicators following \citet[][]{Zechmeister2017}, H$_\alpha $ line index, differential line width (dLW), and chromatic index (CRX).
The HARPS RVs clearly exhibit a significant peak at 35.6\,d, which has no counterpart in the activity indices.
}

\label{gls_GJ143} 
\end{figure}

\begin{figure}[tp]
\begin{center}$
\begin{array}{cc} 

\includegraphics[width=9cm]{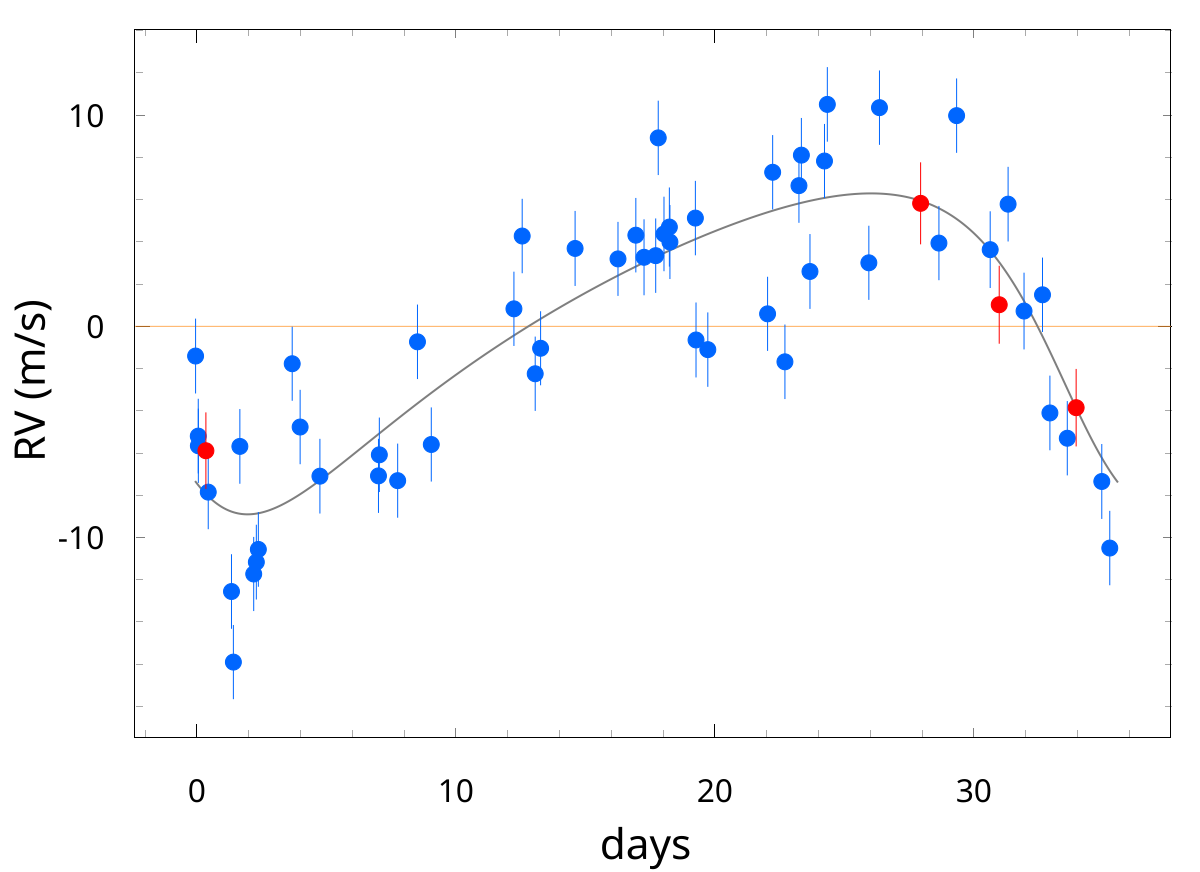} 
\put(-215,160){GJ\,143 b}
\\

\end{array} $
\end{center}

\caption{HARPS RV data  of GJ\,143 phased to the best-fit period of  
$P_{\rm b}$ = 35.589\,d. Blue data points show the pre-fiber upgrade data,
while  red data points represent post-fiber upgrade data. After fitting for 
the mutual RV offset, these data sets are consistent with each other. 
The orbital solution is consistent with the TESS transit event.
The RV uncertainties include the estimated RV jitter, which was quadratically added to the error budget.
}

\label{phase_GJ143} 
\end{figure}

\subsection{RV modeling}

To derive the best-fit model of the RV data, we adopted a maximum likelihood estimator (MLE) scheme 
coupled with a  Nelder-Mead algorithm \citep[][]{NelderMead}, which optimizes the
likelihood function ($-\ln \mathcal{L}$) of a Keplerian or an N-body model.
We modeled the standard the RV curve parameters such as semi-amplitude $K$, period $P$, 
eccentricity $e,$ and arguments of periastron $\omega$. 
Instead of the time of periastron passage $t_p$ , we modeled the 
planetary mean anomalies $M$ for a given epoch. This is a more convenient parameter
in the case of an N-body model, which yields osculating orbital parameters of a given epoch. 
In our study the GJ\,143, and HD\,23472 mean anomalies were defined at the first 
HARPS observational epoch for each data set. 
We assumed edge-on systems, that is, the orbital inclination $i$ was fixed to 90$^\circ$, 
while the longitude of the node $\Omega$ was fixed to 0$^\circ$.
In addition to the fitted parameters, we followed the method of \citet{Baluev2009}
and added the unknown RV variance (RV jitter) as an additional parameter. 

We analyzed the parameter distributions and estimated parameter uncertainties by coupling our 
MLE fitting code to a Markov chain Monte Carlo 
(MCMC) sampling scheme using the $emcee$ sampler \citep{emcee}. 
More details about our fitting tools, including a GUI interface, can be found in 
\url{https://github.com/3fon3fonov/trifon} (Trifonov et al. in prep.),

\begin{figure*}[tp]
\begin{center}$
\begin{array}{cc} 

\includegraphics[width=16cm]{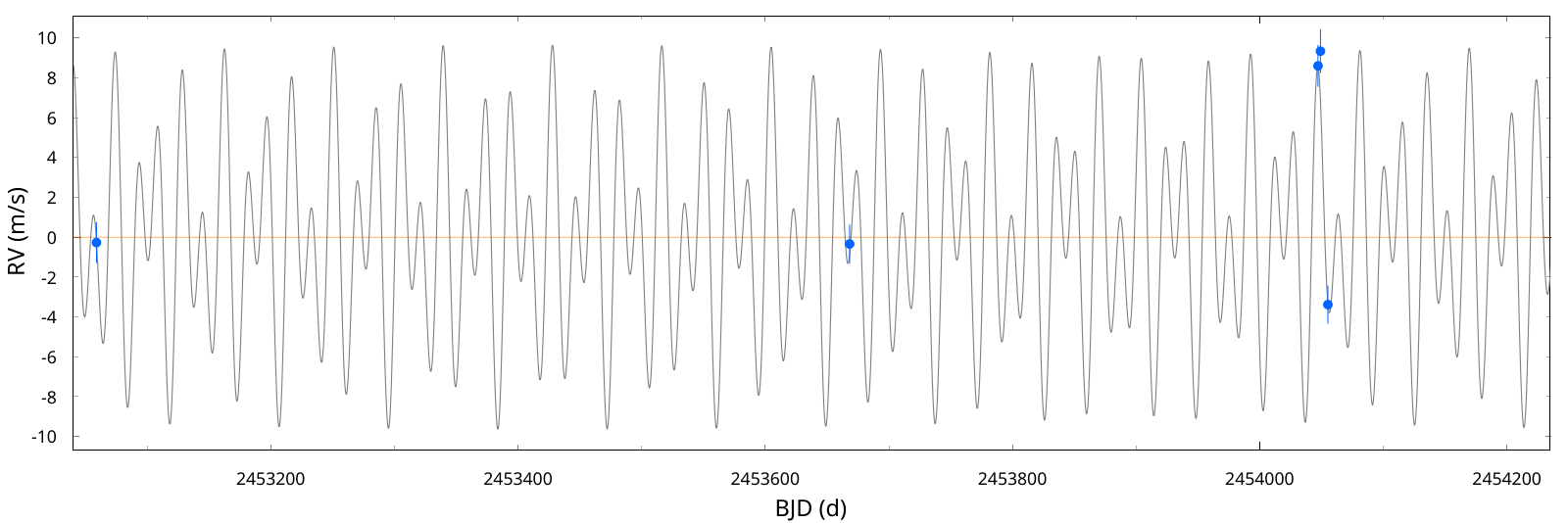}\\
\includegraphics[width=16cm]{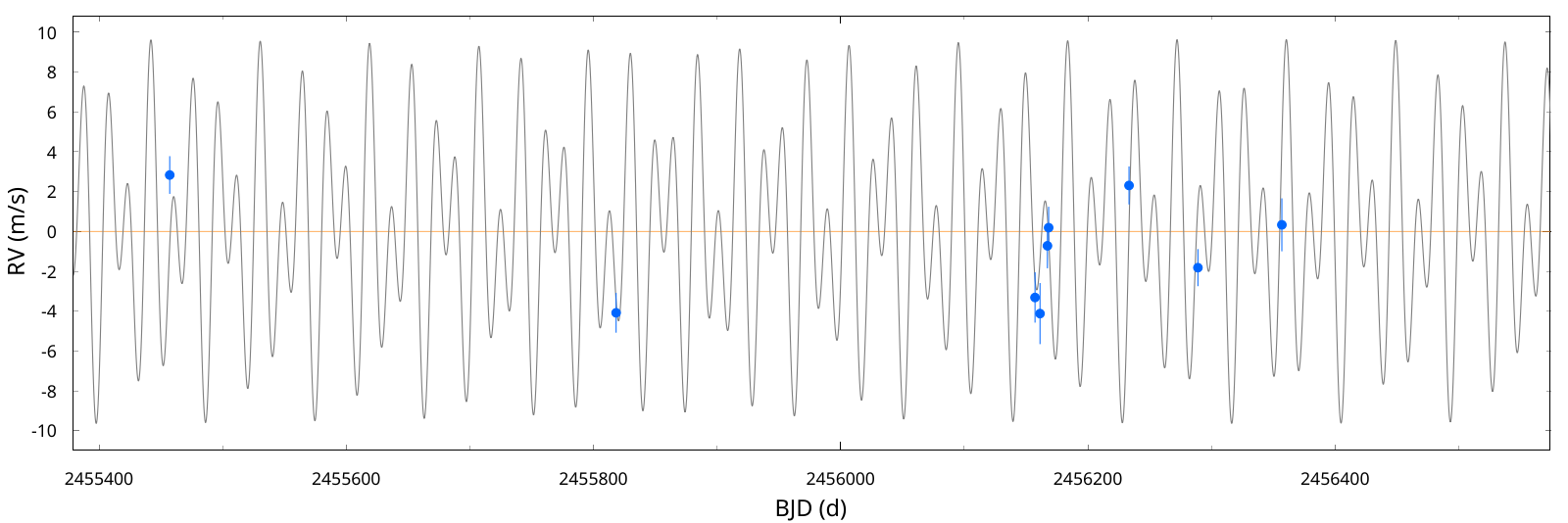}\\

\end{array} $
\end{center}

\caption{Doppler data time-series of HD\,23472 obtained with HARPS. 
The small number of data points (14) by itself is not sufficient to discover the planets,
but taking advantage of the {\em \textup{a priory}} knowledge of the TESS transit ephemerids, it enables
constructing a two-planet model, which is in excellent agreement with the RV data (see Table~\ref{table:orb_param} for details). 
As in Fig.~\ref{phase_GJ143}, the RV uncertainties include the RV jitter. 
}

\label{HD23472_ts} 
\end{figure*}

\begin{figure*}[tp]
\begin{center}$
\begin{array}{cc} 

\includegraphics[width=6cm]{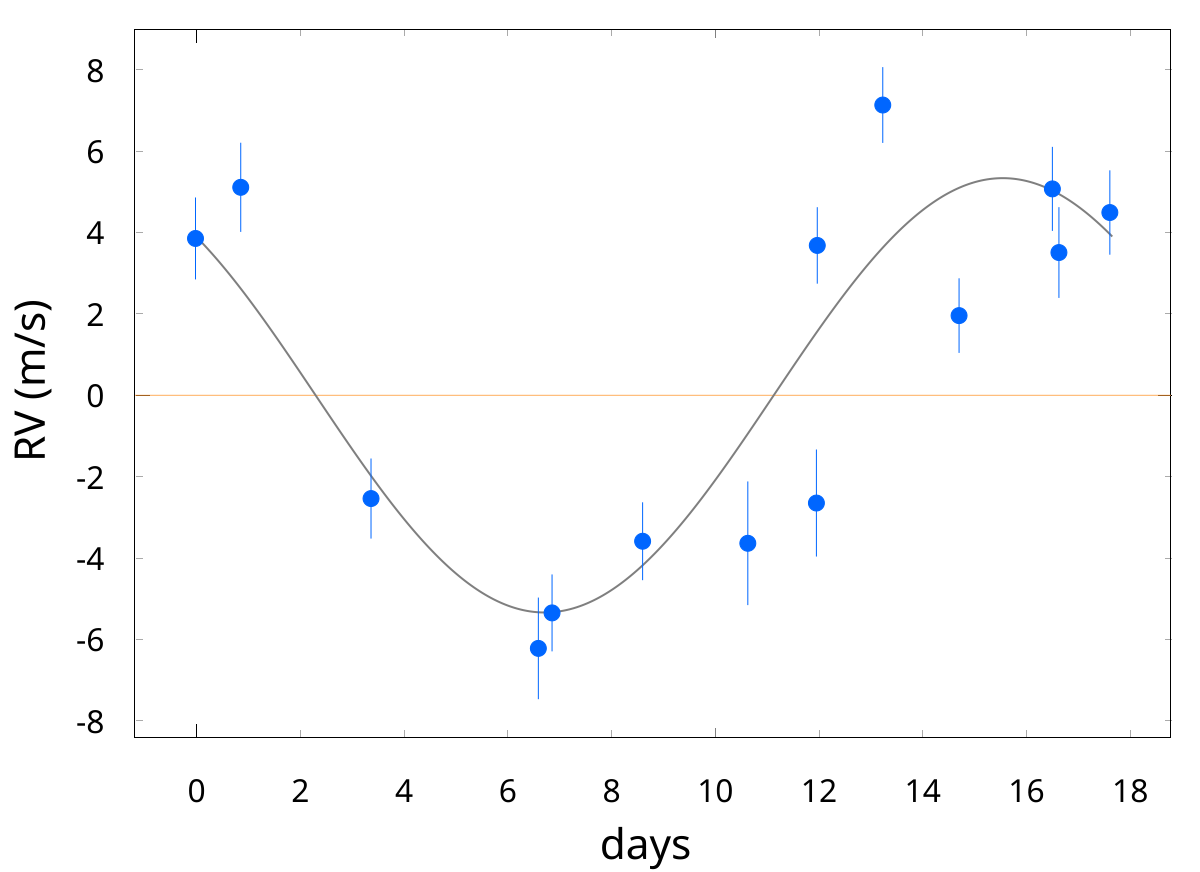} 
\put(-140,105){\tiny HD\,23472 b}
\includegraphics[width=6cm]{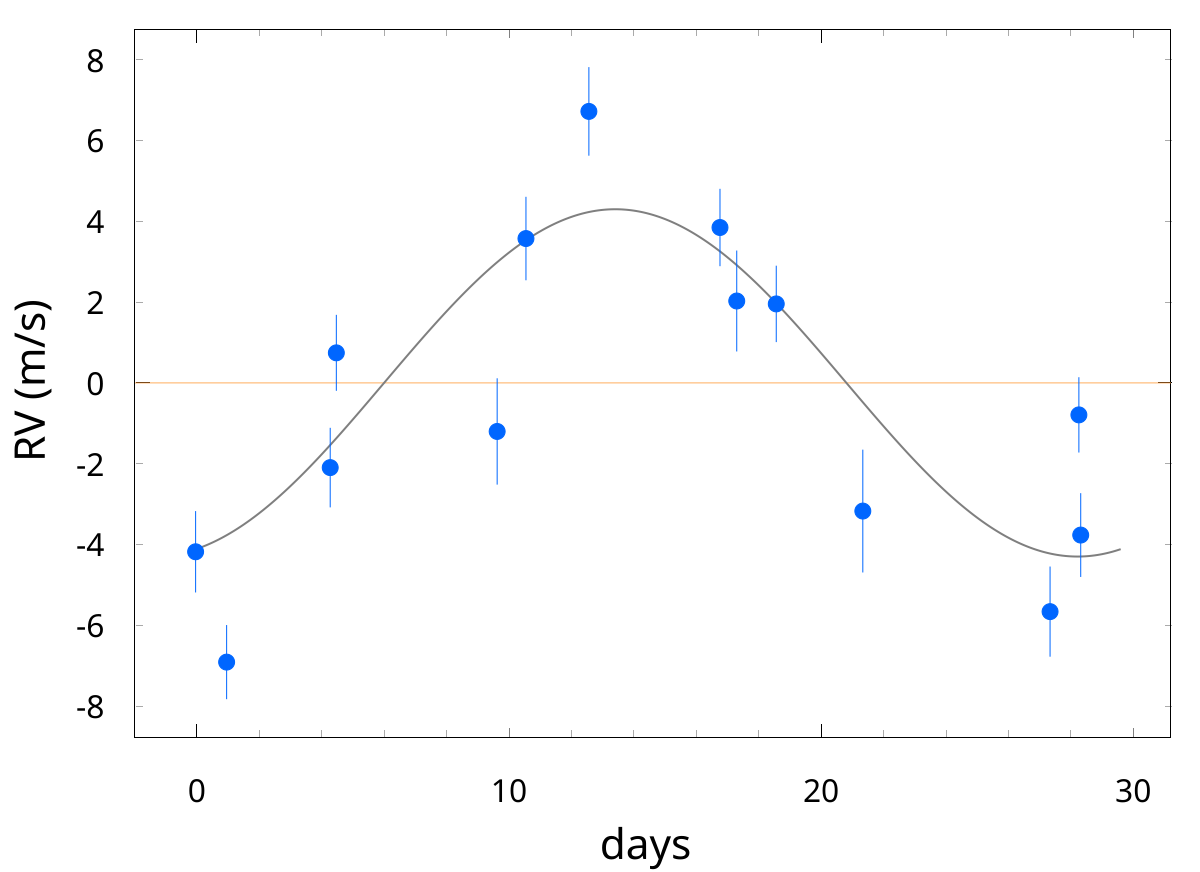} 
\put(-140,105){\tiny HD\,23472 c}
\includegraphics[width=6cm]{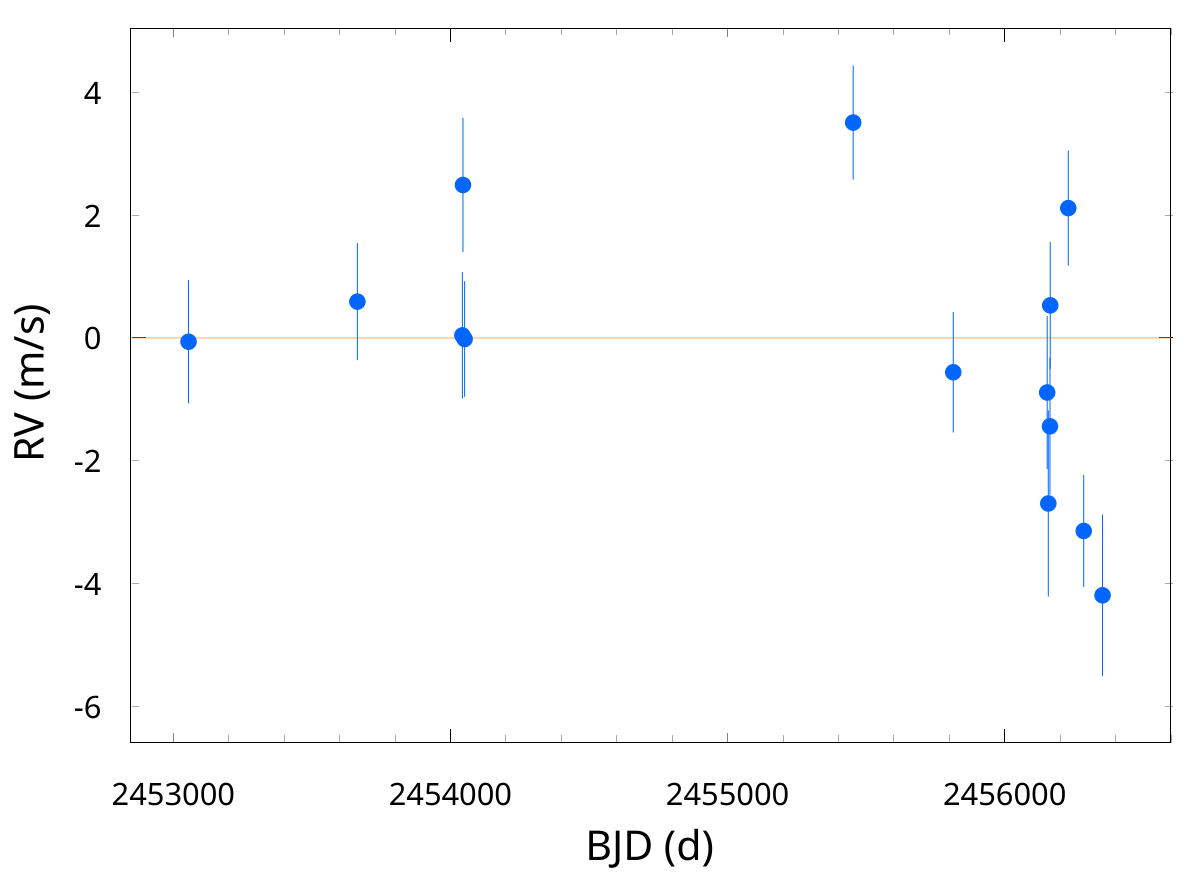}
\put(-140,105){\tiny residuals}
\\


\end{array} $
\end{center}

\caption{Same model and data for HD\,23472 as in Fig.~\ref{HD23472_ts}, but folded to the 
best circular periods for each planet. Shown from left to right are the phase-folded RVs pertaining to 
planets HD\,23472 b and c and the residuals from the model.
}

\label{HD23472_phase} 
\end{figure*}

\section{Results}
\label{Sec3}
 
\subsection{GJ\,143}
\label{Sec3.1} 

The official TESS Sectors 1 and 2 data release has not provided a transit period 
of the GJ\,143 candidate companion because only one single but significant transit  (transit depth = 929 $\pm$ 56\,ppm, R$_{\rm b}$ = 2.56 $\pm$ 0.32\,$R_\oplus$) of 
GJ\,143 has been observed between 22 July to 21 September for this target.
From the transit alone, we therefore have no prior knowledge of the period, but the RV 
data together with the mid-time of the detected transit might in principle reveal the planetary parameters.

Figure~\ref{gls_GJ143} shows a generalized Lomb-Scargle periodogram \citep[GLS;][]{Zechmeister2009} 
of the SERVAL time-series products of the GJ\,143 HARPS spectra.
The RV measurements exhibit a strong peak at 35.62\,d with a 
significant GLS power of 0.6721 that indicates our planetary candidate.
There is no significant power at this period in the HARPS-SERVAL activity indicators. 
It is worth noting that the measurements of the differential line-width activity indicator 
display significant excess power at very small frequencies and at a period of $\approx 400$\,d,
the latter confirmed by the H$_\alpha$ index. This may mean that GJ\,143 is a somewhat active star.

Because the indication of a planet-induced signal is strong, we applied a full Keplerian model
to the RV data. However, while we were sampling around the best-fit model, we rejected all 
the confident MCMC samples whose configurations would not lead to a transit event at BJD = 2458350.312
with a tolerance of $\pm$ 2h.
We converged to a best-fit solution consistent with a planet with a mass of 30.6 $M_\oplus$ (almost twice
the mass of Neptune) orbiting with a period of $P_{\rm b}$ = 35.6\,d on a 
moderately eccentric orbit with $e_{\rm b}$ = 0.33.
Orbital parameter estimates and MCMC uncertainties for GJ\,143 b 
are provided in Table~\ref{table:orb_param}.
The best-fit solution and its confidence region are fully consistent with the 
transit event reported by TESS. 

\begin{table}[tbp]

\centering  

\caption{Keplerian parameters of GJ\,143 b and a two-planet 
circular configuration of HD\,23472 based on HARPS archival RV measurements.}  

\resizebox{0.70\textheight}{!}{\begin{minipage}{\textwidth}

\label{table:orb_param}      
  
\begin{tabular}{l r r r r r r r r r r r}     

\hline \hline \noalign{\vskip 0.7mm}  

\makebox[0.1\textwidth][l]{\hspace{24 mm}One-planet\hspace{11 mm}Two-planet Keplerian} \\
\makebox[0.1\textwidth][l]{\hspace{24 mm}Keplerian \hspace{11 mm} with circular orbits} \\

\cline{2-2}\cline{4-5} \noalign{\vskip 0.9mm} 

Orb. param. & GJ\,143 b && HD\,23472 b  &  HD\,23472~c      \\     
\hline\noalign{\vskip 0.5mm} 
\noalign{\vskip 0.9mm}

$K$  [m\,s$^{-1}$]       &  7.60$_{-0.65}^{+0.64}$       && 5.33$_{-4.20}^{+0.67}$     &  4.29$_{-3.44}^{+0.26}$        \\  \noalign{\vskip 0.9mm}
$P$ [d]                  & 35.589$_{-0.005}^{+0.006}$    && 17.667$_{-0.095}^{+0.142}$  &  29.625$_{-0.171}^{+0.224}$      \\  \noalign{\vskip 0.9mm} 
$e$                      &  0.325$_{-0.079}^{+0.079}$      && 0.0  (fixed) &  0.0 (fixed)      \\  \noalign{\vskip 0.9mm}
$\varpi$ [deg]           & 121.8$_{-19.1}^{+19.2}$         && 0.0  (fixed) &  0.0 (fixed)       \\  \noalign{\vskip 0.9mm}
$M$ [deg]                & 10.2$_{-14.4}^{+14.5}$           && 42.9$_{-35.5}^{+260.3}$       &  196.7$_{-79.0}^{+107.3}$           \\  \noalign{\vskip 0.9mm}

$a$ [au]                 & 0.1932$_{-0.0002}^{+0.0002}$ && 0.121$_{-0.001}^{+0.001}$   & 0.170$_{-0.001}^{+0.001}$       \\  \noalign{\vskip 0.9mm}
$m_{\rm p}$  [$M_{\oplus}$]   & 30.63$_{-2.67}^{+2.63}$      && 17.92$_{-14.00}^{+1.41}$   &  17.18$_{-13.77}^{+1.07}$  \\  \noalign{\vskip 3.9mm}

\hline

\end{tabular} 
\end{minipage}}

\end{table}

\subsection{HD\,23472}
\label{Sec3.2}  
 
  The two transiting exoplanet candidates around HD\,23472 inferred from 
  TESS observations have periods of P$_{\rm b}$ = 17.6800 $\pm$ 0.0015\,d and
  P$_{\rm c}$ = 29.8102 $\pm$ 0.0047\,d, respectively. 
  The durations of both transits are $\approx 3\pm$ 0.3\,h,  with transit depths of 663 $\pm$ 48\,ppm
  for the inner and 627 $\pm$ 61\,ppm for the outer planet.
  This leads 
  to estimated planetary radii 
  consistent with sub-Neptune-size objects, R$_{\rm b}$ = 1.872 $\pm$ 1.321\,$R_\oplus$ 
  and R$_{\rm c}$ = 2.149 $\pm$ 0.345\,$R_\oplus$, respectively.
  To measure the most likely masses of the planetary candidates, we modeled the available HARPS 
  data using a two-planet circular Keplerian model, and used the transit information as a prior.
  In this case, the only unconstrained parameters in this model are the RV semi-amplitudes $K_{\rm b,c}$
  (related to the planetary masses) because the transit epoch and 
  periods are well established by the photometry. 
  However, we decided to include the planetary periods and phases in the modeling.
 Our reason for this is the following:
 Over the $\approx 14$ years of RV measurements, small deviations of the oscillating orbit 
 may accumulate into strong deviations from an unperturbed sine-like curve in some phases where RV
 data are obtained.  
 Moreover, although they are precise, the number of TESS transits is limited
 (only two for HD\,23472 c), and thus we do not know the scale of the gravitational 
 interaction in the system,  which may have a strong impact on the orbital perturbation and transit timing variations (TTVs).
  We note that taking the exact period estimates from TESS leads to a very poor fit, and allowing the planetary  periods to be adjusted slightly in the modeling is therefore well justified.
 Our best-fit circular model is shown in Figs.~\ref{HD23472_ts} and~\ref{HD23472_phase}, while 
 Table~\ref{table:orb_param} summarizes its parameters and MCMC uncertainties.
 We find very consistent estimates of HD\,23471 b, whose period is very similar to that found by TESS,
 whereas for HD\,23471 c, the period deviates by about 4 h, which is still within the MCMC-derived errors, however.

\begin{figure}[tp]
\begin{center}$
\begin{array}{cc} 

\includegraphics[width=9cm]{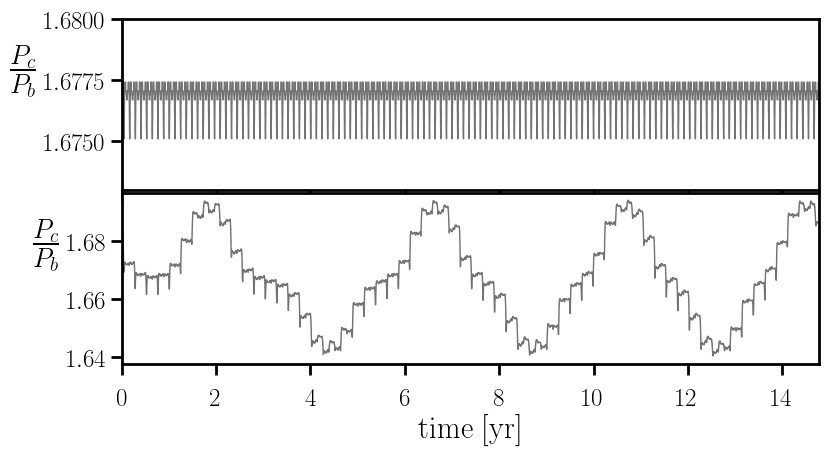} 
\put(-210,125){\tiny a) }
\put(-210,70){\tiny b) }
\\
 
\end{array} $
\end{center}

\caption{Numerical integration of the HD\,23472 system between the 
first HARPS data epoch and the TESS transit event epoch on the inner planet. 
Panel a) shows the evolution of the two-planet circular fit tabulated in 
Table~\ref{table:orb_param}.
Panel b) shows the evolution of a full dynamical model to the 
HARPS data, which is consistent with a moderately eccentric inner planet at start ($e_{\rm b int.}$ = 0.2)
and thus stronger orbital oscillations. 
The mean period ratio in both cases is slightly above 5:3, which may indicate a resonant system.
}

\label{HD23472_p_rat} 
\end{figure}

For completeness, we also tested a full two-planet dynamical model, which led to a marginally better fit. 
 This fit suggests a somewhat eccentric inner planet with $e_{\rm b}$ = 0.2, and 
 the dynamical integration of this configuration revealed 
 a much larger oscillation of the period ratio of P$_{\rm c}$/P$_{\rm b}$ from 
 1.64 to 1.70 with a mean period of 1.67, hence in a potential 5:3 mean motion resonance (MMR).
 Figure~\ref{HD23472_p_rat} shows the N-body evolution of the P$_{\rm c}$/P$_{\rm b}$ 
 of the two-planet circular and the full dynamical models within the baseline of the 
 observations \citep[using SyMBA;][]{Duncan1998}. Clearly, both configurations are dynamically active,
 which shows how challenging the multiplanet modeling of sparse archival data could be because of the unknown magnitude of the dynamical perturbations in the system.
 The full N-body model is more realistic and may to some extent explain 
 the small deviation of the period of HD\,23472 c, but when compared with the TESS results,
 modeling these sparse 14 HARPS RVs with such a complex model is not justified.
 We therefore limit our dynamical analysis of the HD\,23472 system to an initially
 circular solution (i.e., fixed $e_{\rm b,c} = 0, \omega_{\rm b,c} = 0$), which 
 is already a good model. 
 
 We caution that more RV data may be needed to set firm constraints 
 to the orbital parameters and dynamical masses of the HD\,23472 system. This can be effectively done
 by modeling the transit photometry and RVs simultaneously when more data are collected.
 Until then, the ambiguity in the models continues, although 
 the magnitude of the HD\,23472 b and c planetary masses is already determined in the range of 
 a few Earth masses to more likely Neptune-mass planets. This was our 
 scientific goal with this paper.

\section{Conclusions}
\label{Sec4}

We reported the confirmation of two new planetary systems discovered with TESS.
The GJ\,143 system is an interesting case because based on the orbital period of the 
GJ\,143 companion, TESS was able to detect only one single transit, 
and thus no reported period was available.
However, the RV data of GJ\,143 together with the transit constraints from TESS
revealed an eccentric massive Neptune with 30.6$_{-2.6}^{+2.7}$ $M_{\oplus}$ 
and a period of $P_{\rm b}$ = $35.59_{-0.01}^{+0.01}$\,d.
We provided strong arguments for a two-planet system around HD\,23472, 
which is likely composed of two Neptune-mass planets with 17.9$_{-14.0}^{+1.4}$ $M_{\oplus}$ 
and 17.2$_{-13.8}^{+1.1}$ $M_{\oplus}$, respectively. 
The two planets are consistent with a period ratio of 5:3, which may indicate 
a second-order MMR system. 

GJ\,143 would benefit from more photometric data to fully constrain the transit period 
and the system configuration in general. Similarly, HD\,23472 would benefit from 
more RV measurements, which together with the transit photometry could 
lead to a more conclusive dynamical analysis, revealing the  
exact planetary masses and possible resonant motion involved.

We validated the existence of these two TESS planetary system candidates using HARPS archival data taken prior to the TESS observations.
This shows the importance of archival Doppler data in the TESS era.

{\it {\bf Note:} While this paper was in review, we became aware of an independent
analysis of the GJ\,143 system by \citet{Dragomir2019}.
Based on additional TESS Sector 3 photometry, these authors also
validate GJ\,143 b and report  
a possible existence of an additional 
planet candidate (TOI 186.02) with a period of $\sim$7.8 days.
}

\begin{acknowledgements}
JR was funded by the DLR (German space agency) via grant 50\,QG\,1403.
This research has made use of the SIMBAD database, operated at CDS, Strasbourg, France.
This work has made use of data from the European Space Agency (ESA)
mission {\it Gaia} (\url{https://www.cosmos.esa.int/gaia}), processed by
the {\it Gaia} Data Processing and Analysis Consortium (DPAC,
\url{https://www.cosmos.esa.int/web/gaia/dpac/consortium}). Funding
for the DPAC has been provided by national institutions, in particular
the institutions participating in the {\it Gaia} Multilateral Agreement.
This paper includes data collected by the TESS mission. 
Funding for the TESS mission is provided by the NASA Explorer Program.
This research has made use of the Exoplanet Follow-up Observation Program 
website, which is operated by the California Institute of Technology, 
under contract with the National Aeronautics and Space Administration under 
the Exoplanet Exploration Program. 
This paper includes data collected by the TESS mission, which are
publicly available from the Mikulski Archive for Space Telescopes (MAST).
We are extremely grateful to all the people who with the TESS mission and 
light curves make this science possible.


\end{acknowledgements}

\bibliographystyle{aa}

\bibliography{carm_bib}

\clearpage

\end{document}